# Superconductivity of up to 80 K for Tb-123 (TbSr$_2$Cu$_{2.7}$Mo$_{0.3}$O$_{7+\delta}$)

V.P.S. Awana[1,*], H. Kishan[1], E. Takayama-Muromachi[2], T. Watanabe[3], M. Karppinen[3], H. Yamauchi[3], S.K. Malik[4], W.B. Yelon[5], V. Ganesan[6] and A.V. Narlikar[6]

[1]National Physical Laboratory, K.S. Krishnan Marg, New Delhi -110012, India.
[2]National Institute for Materials Science, 1-1, Namiki, 305-0044, Tsukuba, Ibaraki, Japan.
[3]Materials and Structures Laboratory, Tokyo Institute of Technology, 226-8503, Yokohama, Japan.
[4]Tata Institute of Fundamental Research, Homi Bhabha Road, Mumbai 400005, India.
[5]Graduate Center for Materials Research, University of Missouri-Rolla, MO 65409, USA
[6] Inter-University Consortium for DAE Facilities, University Campus, Khandwa Road, Indore-452017, MP India.

The Tb-123 phase has been synthesized in single-phase form with the composition TbSr$_2$Cu$_{2.7}$Mo$_{0.3}$O$_{7+\delta}$ by solid-state reaction route and its phase purity is confirmed by neutron powder diffraction experiments. As-synthesized sample does not show superconductivity down to 5 K. An unusually high antiferromagnetic ordering temperature ($T_N$) of around 7 K is seen for Tb. After 120-atm-O$_2$ post-annealing, bulk superconductivity is achieved in the compound at around 30 K, without any significant effect on $T_N$. To achieve higher oxygen content and higher $T_c$, the as-synthesized sample is subjected to high-pressure oxygenation (HPO) carried out in a closed cell at 5 GPa and 400 °C in the presence of AgO as an excess-oxygen source. This sample exhibited superconductivity onset at around 80 K with a Meissner fraction of larger than 10 % at 5 K. Our observation of superconductivity at 80 K is the highest $T_c$ to-date among the Tb-123 compounds.

PACS No. 74.25Fy, 74.25Ha, 74.72 Jt,

* Corresponding Author:
E-mail: awana@csnpl.ren.nic.in
Fax No. 0091-11-5726938/5726952

## I. INTRODUCTION

Since the discovery of high transition temperature superconductivity (HTSc) in *RE*-123 (REBa$_2$Cu$_3$O$_{7\pm\delta}$, *RE* = rare earth element) compounds, the Ce, Pr and Tb containing *RE*-123 compounds have occupied a unique place [1,2]. In the case of Ce and Tb, the ideal crystal structure is not formed due to tetravalent state of the former and strong intermixing with Ba of the latter [3,4]. The situation for Pr is even more interesting. Though the Pr-123 phase forms, it is not superconducting [5]. The case of non-superconducting Pr-123 with unusually high Néel temperature, $T_N$, of 17 K for Pr has been discussed in detail in literature over a decade [6,7].

For the case of Tb-123, obtaining the phase has not been possible with the composition TbBa$_2$Cu$_3$O$_{7+\delta}$ by any means due to strong intermixing of Tb and Ba at their respective sites [4,8]. To overcome the intermixing problem Ba is replaced with Sr, followed by small amount (~30 %) of Cu in CuO chains being replaced by Mo for stabilization of the structure. There are a couple of reports about the formation of TbSr$_2$Cu$_{2.7}$Mo$_{0.3}$O$_{7+\delta}$ compounds in ideal *RE*-123 structure in which superconductivity is achieved with $T_c$ up to 30-40 K at maximum [9-11]. Our aim was to explore further possibilities of enhancing the $T_c$ in Tb-123. For this very reason we synthesized a single phase TbSr$_2$Cu$_{2.7}$Mo$_{0.3}$O$_{7+\delta}$ sample and subjected it to various oxygen loading procedures. We could achieve superconductivity in this compound with $T_c$ up to 80 K, though with small Meissner volume fraction. To our knowledge this is the first ever synthesized Tb-123 compound with superconductivity at 80 K.

## II. EXPERIMENTATION

The Tb-123 sample with the composition TbSr$_2$Cu$_{2.7}$Mo$_{0.3}$O$_{7+\delta}$ was synthesised through a solid-state reaction route starting from Tb$_4$O$_7$, MoO$_3$, SrO$_2$, and CuO. Calcinations were carried out on mixed powders at 975 $^o$C, 990 $^o$C, and 1000 $^o$C each for 24 hours with intermediate grindings. Finally the sample was annealed in air for 24 hours at 1000 $^0$C. This sample is named as "as-synthesized". A part of the as-synthesized sample was post-annealed in a flow of high-pressure oxygen (120 atm) at 420 $^o$C for 12 hours and 360 $^o$C for 12 hours and subsequently cooled slowly to room temperature. This sample is named as "120-atm-O$_2$ annealed". Another part of the as-synthesized sample (~50 mg) was mixed with 100 mol-% of AgO, sealed in a gold capsule and annealed in a high-pressure apparatus at 5 GPa and 400 $^o$C for 30 minutes. This high-pressure oxygenated sample is named "HPO treated". Oxygen content of the as-synthesized and the 100-atm-$O_2$ annealed sample was determined from coulometric titrations at 7.12 (average Cu valence 2.01) for the former and at 7.15 (average Cu valence 2.04) for the latter.

Magnetization measurements were carried out using a SQUID magnetometer (Quantum Design: MPMS-XL). Neutron diffraction patterns were obtained at room temperature at the University of Missouri Research Reactor Facility, for which the samples were contained in a thin-walled vanadium sample holder. Neutrons of wavelength 1.4783 Å were selected for the diffraction experiments. The diffraction data were recorded from $10^0$ to $100^0$, with $0.05^0$ intervals, using a 5-element position-sensitive array covering $20^0$ at a time.

### III. RESULTS AND DISCUSSION

Room temperature neutron powder diffraction (NPD) pattern for the as-synthesized $TbSr_2Cu_{2.7}Mo_{0.3}O_{7+\delta}$ sample was found nearly free from any impurities (Figure not shown). The same was readily fitted in tetragonal space group *P4/mmm* with lattice parameters *a* =3.83709(6) Å and *c* = 11.57837(1) Å. The *RE*-123 structure can be viewed as $(Ba,Sr)O/CuO_2/RE/CuO_2/(Ba,Sr)O$ slabs interconnected by a sheet of Cu and O atoms together with a possible substituent *M* (here *M* = Mo) with variable composition, $(Cu,M)O_{1\pm\delta}$. The oxygen sites in the $CuO_2$ plane are identified as O(2) and O(3). Oxygen site in the SrO plane is called O(4). The *RE* plane is devoid of any oxygen. The oxygen sites in the $CuO_{1\pm\delta}$ layer are named O(1) (along *b* axis) and O(5) (along *a* axis). In the tetragonal *RE*-123 structure, O(2) and O(3) are indistinguishable and so are O(1) and O(5) since lattice parameters *a* and *b* become identical. The neutron diffraction data on the present Tb-123 sample have been analyzed by the Rietveld refinement procedure using the generalized structural analysis system (GSAS) program. Neutron scattering lengths used are (in units of fm) 0.738 for Tb, 0.702 for Sr, 0.7718 for Cu, 0.6715 for Mo and 5.05 for O. Structural parameters, including atomic coordinates, occupancy and thermal parameters ($U_{iso}$) for different atoms including variously named oxygen atoms are listed in in Table 1. The overall stoichiometry of the compound, obtained from various atom occupancies is close to the nominal cation composition with oxygen content of 7.32. Interestingly this oxygen-content value is higher than being obtained from coulometric titrations (7.12). It is seen from Table 1, that even though the O(2) = O(3), occupancy ($CuO_2$-plane oxygen) is full, the O(4) occupancy is incomplete. The BaO plane lies next to $(Cu,Mo)O_{1+\delta}$ layer and any instability in oxygen atoms in this layer affects directly the O(4).

This instability is clear from the fact that the thermal parameters for O(1) are reasonably higher than those for other oxygen atoms in the structure. Further to get a reasonable fit to the neutron data, we had to split the charge-reservoir oxygen sites to O(1) and O(1b) with two different $x$ positions, see Table 1. In such a situation, when the oxygen in $(Cu,Mo)O_{1+\delta}$ can not be fixed with reasonable thermal parameters and the O(4) is not fully occupied, the overall oxygen content of the compound from NPD data may differ from the coulometric titration value. Conclusively from neutron diffraction data of as-synthesized $TbSr_2Cu_{2.7}Mo_{0.3}O_{7+\delta}$ sample, we can say that the studied compound is single phase in nature and the oxygen atoms in the $(Cu,Mo)O_{1+\delta}$ layer do have some degree of instability. This instability can arise from the fact that lower-valent Cu is partially substituted by higher-valent Mo in this layer.

Figure 1 shows the dc magnetic susceptibility of the as-synthesized sample of Tb-123 in an applied field of 5000 Oe. The sample exhibits paramagnetic behaviour above 7 K and an antiferromagnetic transition of Tb moments below this temperature. The effective paramagnetic moment of Tb was estimated by fitting the magnetic susceptibility data to Curie-Weiss equation, and found to have a value of 9.80 $\mu_B$ per Tb atom. However such an exercise does not necessarily lead to a conclusive value due to presence of crystal-field effects and the unknown contribution to the susceptibility from Cu in the compound. As far as $T_N$ of 7 K is concerned, this is unusually high for Tb in a 123-type compound. For example, a $T_N$ of ~1.8 K is seen for Gd (8 $\mu_B$) in Gd-123 compounds [1,2]. Interestingly, even though the localised magnetic moments of Gd and Tb are comparable, the $T_N$ of the latter is more than 3 times the $T_N$ of the former within the same 123 structure. It has been suggested earlier that the RKKY mechanism itself might not be applicable in the case of Tb-123 and the dipole-dipole interactions might explain the unusually high $T_N$ in Tb-123 [12].

Figure 2 shows the magnetic susceptibility (measured in an applied field of 10 Oe) versus temperature plot for the 120-atm-$O_2$ annealed Tb-123 sample. It is clear from this figure that this sample is superconducting below 37 K. The superconducting volume fraction, determined from the field-cooled magnetization is roughly 20 %. Though this value of superconducting transition temperature, $T_c$, is comparable to that obtained in earlier reports [9-11], the volume fraction is higher in the present case. In general, the result shown in Fig. 3 is not yet comparable to the 90 K superconductivity routinely observed in Y-123. Hence we believe that even 120-atm-$O_2$ annealing has not been able to provide sufficient number of mobile hole-carriers in $CuO_2$ planes to achieve the 90-K superconductivity.

To reach higher $T_c$ values, the as-synthesized sample was mixed with AgO as an excess-oxygen source and HPO annealed at 5 GPa and 400 $^o$C. The magnetic susceptibility versus temperature plot for the HPO-treated Tb-123 sample is shown in Fig. 3. The applied field was 10 Oe. It is seen from Fig. 3 that the HPO-treated sample is superconducting up to around 80 K. The superconducting volume fraction is calculated at 14.5 % from the ZFC curve and at 7.4 % from the FC curve, giving an average value of 10.9 %. The value achieved for the superconducting transition temperature along with the sufficient volume fraction is the highest ever reported for a Tb-123 compound. The XRD pattern for the HPO-treated Tb-123 sample is shown in Fig. 4.

Figure 5 shows the resistance ($R$) versus temperature plot for the HPO-treated Tb-123 sample. The $T_c^{onset}$ is seen around 80 K, but $R = 0$ state is not achieved down to 5 K. The resistive $T_c^{onset}$ value coincides with the diamagnetic transition temperature value (see Fig. 4). This shows that presently observed superconductivity at 80 K is observed by both magnetic and electrical transport measurements. Worth mentioning is the fact, that the normal state (above $T_c^{onset}$) resistivity versus temperature behaviour is semiconducting in nature, indicating the under-doped state of the sample. Perhaps this is the reason why we did not reach the $T_c \approx 90$ K.

In conclusion, we have been able to achieve superconductivity with the transition temperature of up to 80 K in a Tb-123 ($TbSr_2Cu_{2.7}Mo_{0.3}O_{7+\delta}$) compound by loading the phase with oxygen by means of high-pressure-oxygenation (HPO) annealing. To our knowledge, this presently reached $T_c$ value of 80 K for Tb-123 is the highest ever reported.

**Table 1.** Refined structural parameters for the as-synthesized TbSr$_2$Cu$_{2.7}$Mo$_{0.3}$O$_{7+\delta}$ sample ($\delta$ = 0.12 from coulometric titration), including the atomic coordinates, occupancies, and thermal parameters (U$_{iso}$). Space group $P4/mmm$, lattice parameters are $a$ =3.83709(6) Å and $c$ = 11.57837(1) Å. The uncertainty in occupancies and coordinates is only after 3 decimal points.

| Atom | Occupancy | x | y | z | U$_{iso}$ (Å$^2$) |
|---|---|---|---|---|---|
| Tb | 1.000 | 0.500 | 0.500 | 0.500 | 0.62(1) |
| Sr | 1.000 x2 | 0.500 | 0.500 | 0.1966 | 2.40(3) |
| Cu(1) | 0.700 | 0.000 | 0.000 | 0.000 | 4.54(2) |
| Cu(2) | 1.000 x 2 | 0.000 | 0.000 | 0.3544 | 1.12(2) |
| O(1) | 0.2361 x 4 | 0.2463 | 0.500 | 0.000 | 8.61(5) |
| O(2) | 1.000 x 4 | 0.500 | 0.000 | 0.3732 | 1.35(2) |
| O(4) | 0.9405 x 2 | 0.000 | 0.000 | 0.1590 | 2.77(1) |
| Mo@Cu1 | 0.300 | 0.000 | 0.000 | 0.000 | 3.43(3) |
| Mo@Cu2 | 0.000 | 0.000 | 0.000 | 0.3541 | 2.54(3) |
| O(1b) | 0.1247 x 4 | 0.0760 | 0.500 | 0.000 | 6.35(4) |

**FIGURE CAPTIONS**

Fig. 1. Magnetization of the as-synthesized Tb-123 sample.

Fig. 2. Magnetization of the 120-atm-O$_2$ annealed Tb-123 sample.

Fig. 3. Magnetization of the HPO-treated Tb-123 sample.

Fig. 4. XRD patterns for (a) the as-synthesized and (b) the HPO-treated Tb-123 sample.

Fig. 5. $R$ vs $T$ for the HPO-treated Tb-123 sample.

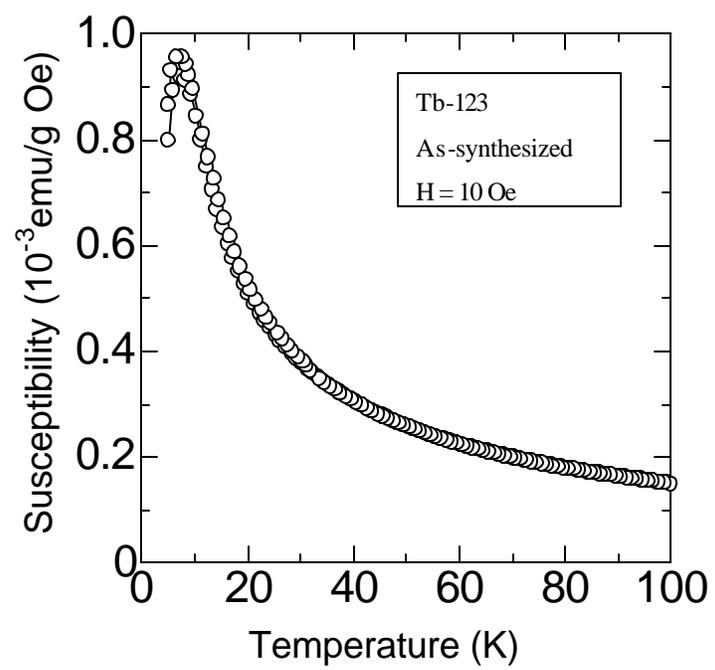

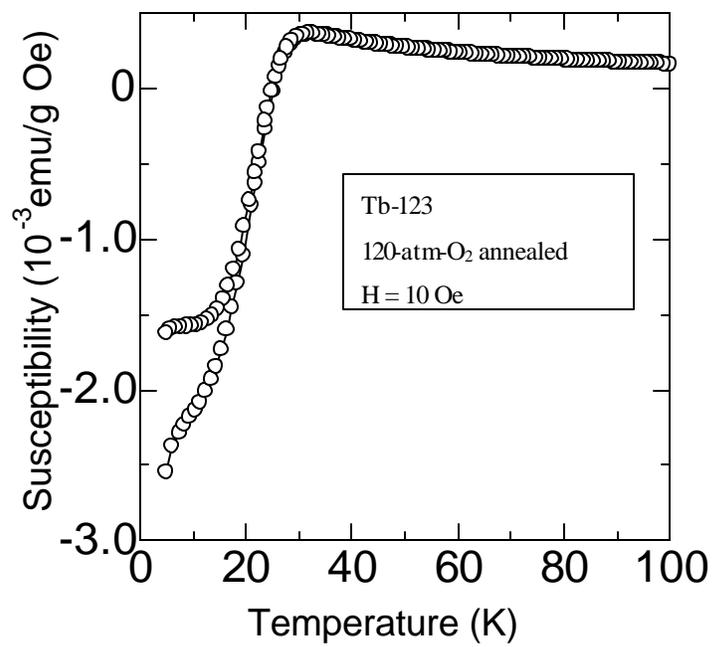

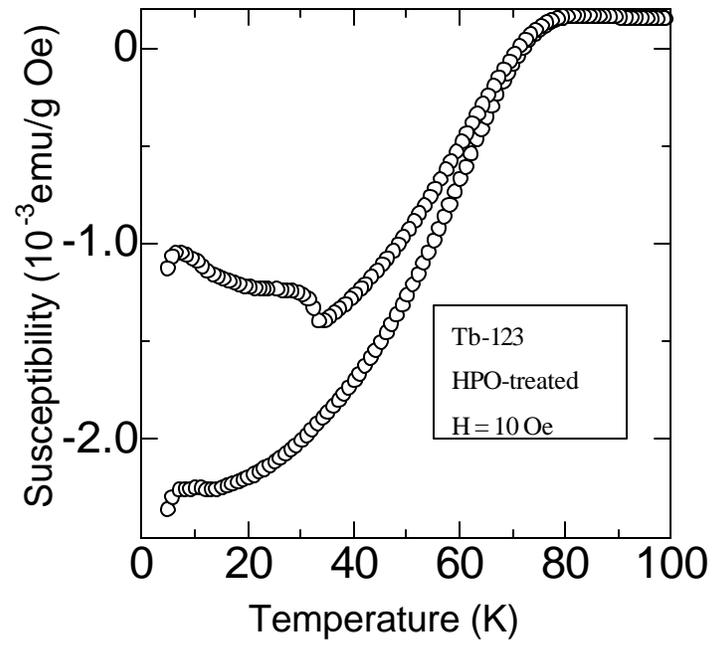

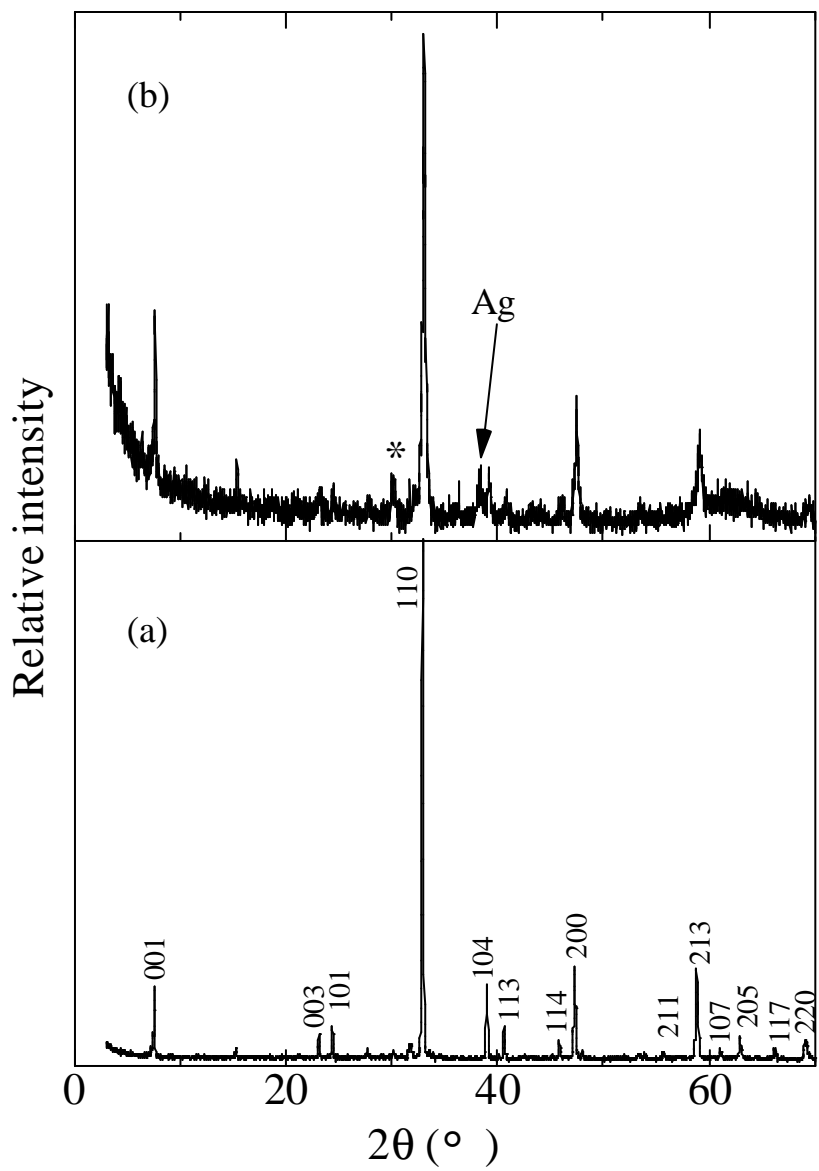

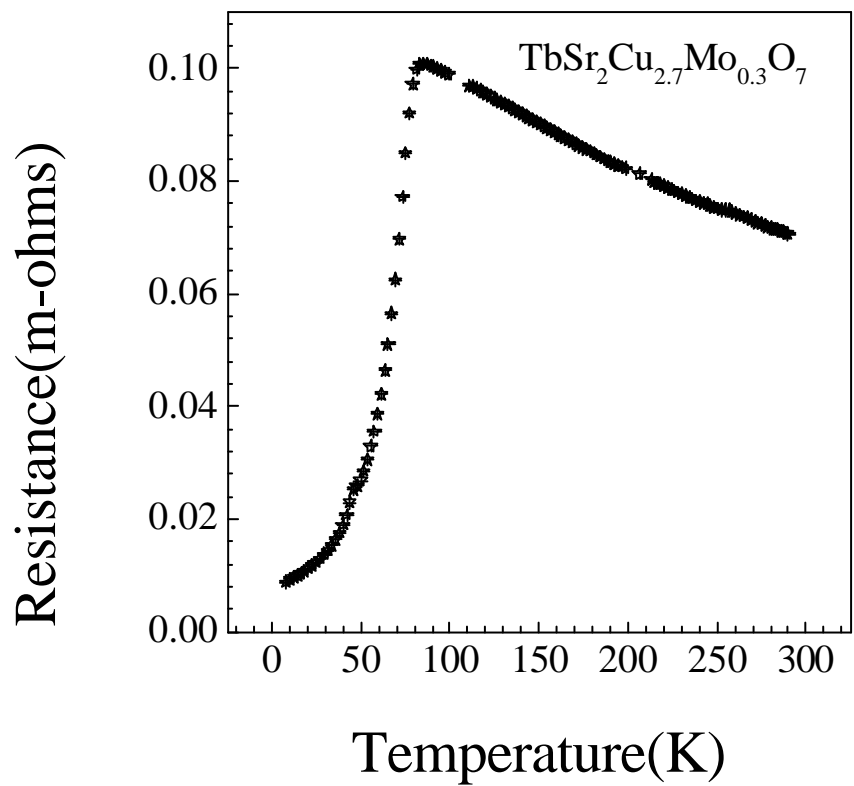